\documentclass[12pt]{iopart}

\usepackage{epsf}
\usepackage{graphicx}

\begin{document}

\title[Measurement of
eddy currents in a liquid metal]{Measurement of
the spatio-temporal distribution of
harmonic and transient eddy currents in a liquid metal}

\author{J Forbriger, V Galindo,  G Gerbeth and F Stefani}

\address{Forschungszentrum Dresden-Rossendorf, P.O. Box 510119,
D-01314 Dresden, Germany}
\ead{F.Stefani@fzd.de}
\begin{abstract}
Harmonic and transient eddy currents in a
liquid metal
positioned above an excitation coil are determined by
measuring the voltage drop in a simple potential probe.
The resulting spatio-temporal eddy current field
is compared with the corresponding
analytical expressions for a conducting half-space.
Further, a deformation of the eddy current distribution
due to a non-conducting torus immersed into
the liquid metal
is measured and compared with numerical results.
The method can be generalized to arbitrary
geometries, and might help to validate numerical models
for non-destructive testing and magnetic inductance tomography.
\end{abstract}

\vspace{2pc}
%\submitto{\MST}
\submitto{\MST}
\maketitle

\section{Introduction}
Eddy currents play a crucial role in
magnetic induction tomography (MIT) \cite{GRIFFITH}, in
non-destructive testing \cite{BLITZ}, and in
different versions of contactless flow measurements
for liquid metals as, for example, in
contactless
inductive flow tomography  (CIFT) \cite{MST,CIFT},
in induction flow-metering based on phase shift measurements \cite{PRIEDE},
and in Lorentz force velocimetry \cite{THESS}.
Usually, the eddy currents in the bulk material
are indirectly inferred from
non-invasive measurements
of external induced magnetic fields.
The  final inverse problem
of inferring the material inhomogeneities or
flow velocities  from the
induced magnetic fields is still a formidable task.

In many cases, it would be desirable
to have independent
eddy current measurements in order to validate
numerical solvers for the forward and the
inverse problem. For evident reasons, a detailed
eddy current
measurement in solid materials is hardly possible.
Up to present, only some eddy current measurements
at the surface of thin solid aluminum plates were
reported \cite{FUJIWARA} and served for the
validation of a numerical benchmark problem.

In contrast to solid metals, liquid metals
are perfectly suited for a full scan of the eddy current
distribution in the bulk of the conducting material.
Actually, various liquid metals (mercury \cite{FOERSTER}, 
Wood's metal 
\cite{BLITZALAGOA}) have been used already in eddy current
testing since they allow an easy defect modelling.
However, to the best of our knowledge, in this
paper  we present the first spatio-temporal
measurements of eddy currents in a liquid metal alloy
(GaInSn)
both for the case of
harmonic excitation and for the case of a pulsed (transient)
excitation.
Whereas harmonic excitations have been used for long in
non-destructive testing, the transient excitation
has achieved much attention
only during the last
ten years. Nowadays it is used for non-destructive
testing in many areas,
including the detection of corrosion and cracks
in aging
aircrafts fleets \cite{SMITH}.

In order to start with an accurate reference model,
we determine harmonic and transient eddy currents
in a cylindrical vessel which is large compared to the
radius and the distance of the exciting cylindrical coil.
This problem can be reasonably approximated by
the eddy current problem in a
conducting half-space for which analytical solutions
are known, both in the harmonic and
in the transient case \cite{EDDYBUCH}.

Another measurement is carried out for the
more complicated situation that a non-conducting torus
with quadratic cross-section
is immersed
into the liquid metal. These measurements are
then compared with
numerical results from the commercial FEM software OPERA.

The paper closes with a summary and some conclusions.

\section{Theory}
\subsection{Analytical models}
Consider a conducting half-space with conductivity $\sigma$
and a coil of radius $R$ located at a
distance $a$ to the boundary of the half-space (figure 1).
Independent of the time-dependence of the exciting current,
the eddy currents in the half-space will only have an axisymmetric
azimuthal component $j_{\phi}(r,z)$.

\begin{figure}[ht]
\centering
\includegraphics{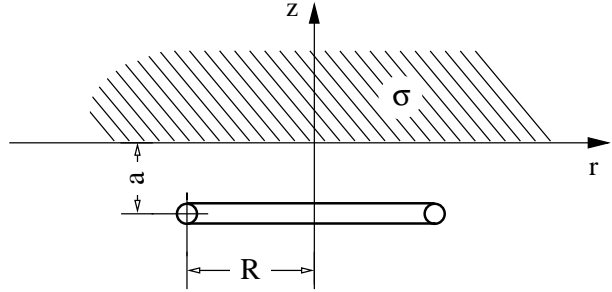}
\caption{Geometry of the simplified analytical model to determine harmonic and
transient eddy currents in the conducting half-space.}
\label{fig1}
\end{figure}

First we consider a harmonic current $I(t)=I_0 \sin(\omega t)$ in the coil
with amplitude $I_0$ and angular frequency $\omega=2 \pi f$.
In \cite{EDDYBUCH}, the  distribution of the eddy
current density $j_{\phi}(z,r,t)$
in the conducting half-space was shown to have the form
\begin{eqnarray}
j_{\phi}(z,r,t)=-i \mu_0 \sigma \omega  R I(t) \int_0^{\infty}
e^{\displaystyle -k a} J_1(kR) J_1(kr) \frac{e^{\displaystyle -q z}}{k+q}
k \; dk \; ,
\end{eqnarray}
where the abbreviation
\begin{eqnarray}
q=\sqrt{k^2+i \mu_0 \sigma \omega} \; .
\end{eqnarray}
is used. $J_1$ is the Bessel function of
order 1 and $\mu_0$ is
the magnetic permeability of the
free space (throughout the paper we assume the
relative magnetic permeability $\mu_{rel}$ to be equal to 1).

For each point $r,z$ the integral (1) can be easily
evaluated. Typically, we have used discretizations
with 50-200 points in the $k$ space, and we have
carefully checked the convergence of the integration
procedure.

Now let us assume that the current in the excitation 
coil is suddenly switched
on so that it corresponds to a step function
$I(t)=I_0 \Theta(t)$.
Quite similar to the harmonic case, the induced transient
eddy current
density $j_{\phi}$ can be represented as an integral over the wave
number $k$:
\begin{eqnarray}
j_{\phi}(z,r,t)=-\mu_0 \sigma  R I_0 \int_0^{\infty}
e^{\displaystyle -k a} J_1(kR) J_1(kr) L(k,z,t) k dk \; ,
\end{eqnarray}
with $J_1$ being again the Bessel function of
order 1 and $L(k,z,t)$ being
defined as
\begin{eqnarray}
L(k,z,t)&=& 
 \left[ \frac{1}{\sqrt{\pi t}} e^{\displaystyle -\frac{\mu_0
\sigma z^2}{4t}}  - \frac{k}{\sqrt{\mu_0 \sigma}}
e^{\displaystyle k|z|+\frac{k^2 t}{\mu_0 \sigma}    }
\mbox{erfc}\left( \frac{k \sqrt{t}}{\sqrt{\mu_0 \sigma}}+
\frac{\sqrt{\mu_0 \sigma} z}{\sqrt{4t}}  \right) \right] \nonumber \\ 
&&\times \frac{e^{\displaystyle -\frac{k^2 t}{\mu_0 \sigma}}}{\sqrt{\mu_0 \sigma}} \; .
\end{eqnarray}

The results of equation (1) for the harmonic case and of equations 
(3,4) for the
transient case will be visualized in the
following section where we will compare them with the
measured current distribution.

\subsection{Finite-element models}

In addition to the analytical expressions (1) and (3), we
also
determine the eddy current by the commercial FEM package OPERA. 
OPERA is a finite element analysis software for time varying
electromagnetic fields. It includes a module for solving
eddy current problems in which the exciting currents can vary
sinusoidally or in another predetermined way in time.

\subsection{The measurement}

The measurement principle for the eddy currents 
relies on Ohm's law in non-moving 
conductors which connects the electric current density $\bf j$ with the
electric field $\bf E$ via ${\bf j}=\sigma {\bf E}$.
The  measured voltage $U_{12}$ between two points $P_1$ and $P_2$ can be
expressed by the line integral over the electric field  
$U_{12}=\int_{P_1}^{P_2} {\bf E}\cdot d{\bf s}$.
The only current component that appears in our particular problem
is the azimuthal one which can be determined by the voltage 
between two electrodes whose difference vector points in azimuthal 
direction. At every instant $t$ and every position $r,z$,  this
azimuthal current density  $j_{\phi}(z,r,t)$ can be approximated
by 
\begin{eqnarray}
j_{\phi}(z,r,t)=\frac{U_{12}(z,r,t) \sigma}{d}
\label{u_i_eqn}
\end{eqnarray}
where $d$ stands for the distance between the electrodes. 
The value of $d$ chosen in the experiment
results as a compromise between maximizing the measurable 
voltage and minimizing the inaccuracies which 
appear in particular for small
radii $r$.

\section{Experimental set-up}

\begin{figure}[ht]
\centering
\includegraphics[width=14cm]{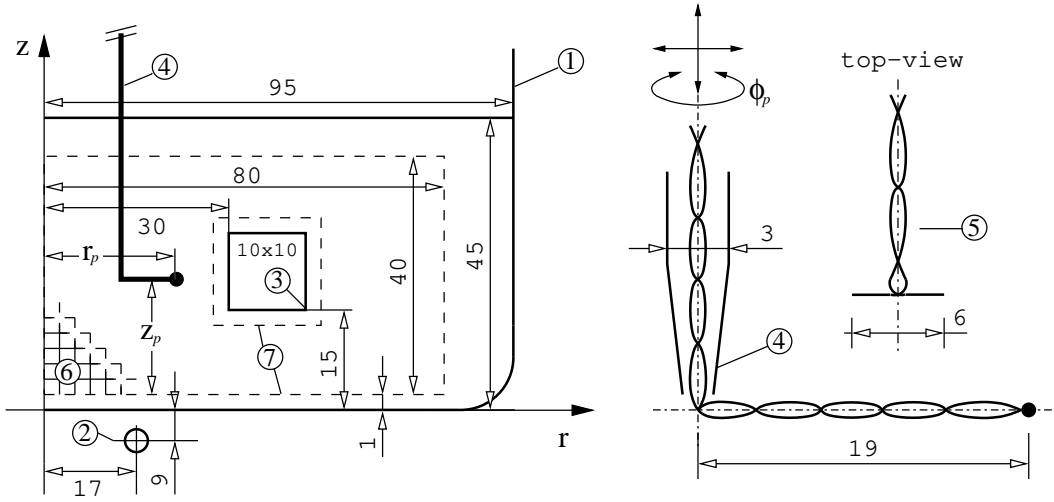}
\caption{Experimental set-up (not to scale). (1) Glass vessel, (2) excitation coil, (3)
PVC ring (only in the second part of the experiment), (4) scanning potential probe,
(5) probe in detail, (6) scan-raster, (7) scan-region. All given dimensions are in mm.}
\label{set_up_fig}
\end{figure}

\begin{figure}[ht]
\centering
\includegraphics{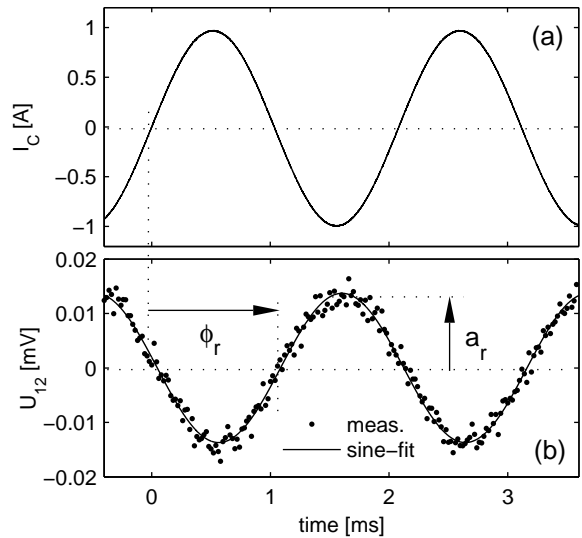}
\caption{Sample-measurement for harmonic excitation at $r_p=20$ mm, $z_p=22.5$ mm.
The coil current $I_c$ (a) has an amplitude of 1 A and a frequency of 480 Hz.
To determine amplitude and phase-angle of the response-signal $U_{12}(t)$,
a sine-function is fitted (least-squares-method) to the data. The
resulting parameters $a_r(r_p,z_p), \phi_r(r_p,z_p)$ are the basis for spatial amplitude
and phase plots.}
\label{i_u_harm_fig}
\end{figure}

\begin{figure}[ht]
\centering
\includegraphics{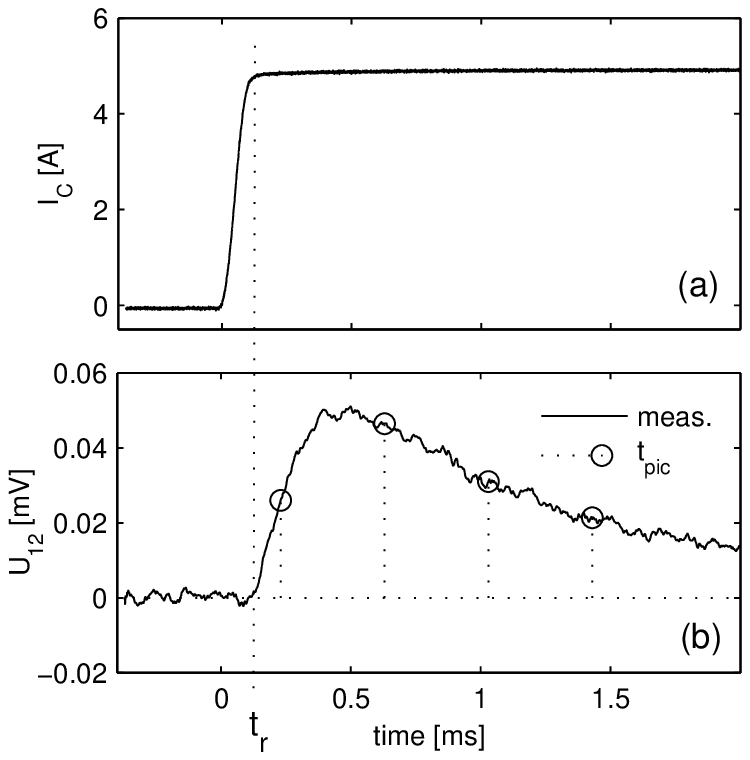}
\caption{Sample-measurement for pulsed excitation at $r_p=20$ mm, $z_p=22.5$ mm. 
The stem-plots in (b) indicate the instants for which the eddy
current distribution will be shown in detail.}
\label{i_u_trans_fig}
\end{figure}

% ################################
% medium
The experimental set-up is shown in \fref{set_up_fig}.
The central part is a  cylindrical glass vessel (1)
filled with the eutectic
alloy Ga$^{67}$In$^{20.5}$Sn$^{12.5}$, which has the advantage 
of being liquid down to temperatures of
about $10^\circ C$. The physical properties of GaInSn at 25 $^{\circ}$C are: density
$\rho=6.36 \times 10^3$ kg/m$^3$, kinematic viscosity $\nu=3.40\times
10^{-7}$ m$^2$/s, electrical conductivity $\sigma=3.27\times 10^6$
($\Omega$ m)$^{-1}$.
The non-conducting ring (3) is immersed into the liquid only 
in the second part of the experiment and is held by one stainless 
steel rod ($d=3$ mm, $\sigma=1.3 \times 10^6$ S/m) which we assume to be
non-essential for the current distribution.

% ################################
% excitation
The external magnetic field is generated by a circular coil 
(2), centered below the vessel (1). This coil consists of 63 turns of a
0.6 mm Cu wire and is wound in such a way
that the coil has a nearly circular cross-section. The coil is fed by
a precise U-I-converter (HERO PA2024C) controlled by an arbitrary-waveform 
generator (AD-win-Pro).
In the case of harmonic excitation, a current of 1 A amplitude and
a frequency of 480 Hz was applied to the coil (see \fref{i_u_harm_fig}a).
In the case of transient excitation,
the limited bandwidth of the amplifier requires that the
pulse-edges have to be approximated by a continuous function.
For this purpose, the 1st
quarter-period of $I(t)=I_0(\sin(2 \pi f t))^2$
with $I_0=5$ A and $f=2$ kHz was chosen. The resulting
rise time (see \fref{i_u_trans_fig}) of $t_r=125$ $\mu$s is small 
compared to the typical diffusion time of the magnetic field 
$t_{diff}=\mu_0 \sigma R^2=1.23$ ms.
The pulse duration was set to $t_{on}=12$ ms, to
avoid interactions between the eddy currents resulting from the
rising and the falling
edge. The pause between the pulses was set to
$t_{off} = 100$ ms. This rather long time
is necessary to avoid a significant heating of the coil.

% ################################
% detection
The spatio-temporal distribution of the induced eddy 
current is determined by pointwise measuring the voltage drop $U_{12}(t)$ 
between the blank ends of two coated wires (0.1 mm Cu-wire)
which are twisted to avoid inductive pick-up (r.h.s. of \fref{set_up_fig}).
The connecting line between the points $P_1, P_2$ points in azimuthal 
direction. The wires are guided through a thin glass pipette (4)
into the medium. In the top-view of the probe (5)
the electrode distance $d=5.8$ mm is visible. 
After differential amplification ($g_d=+60$ dB) and low-pass filtering ($f_c=200$ kHz)
the signal is routed directly into a digital storage oscilloscope (Tektronix TDS3034B),
which records the response with a sample-rate of 2.5 Megasamples/s over 
10 kilosamples. Due to the
signal decrease with respect to the coil-probe-distance, the vertical-sensitivity
of the oscilloscope is adapted to exploit its dynamic range as good as possible.
The entire signal path is realized as a dc-measurement to get correct transients.
To specify the exact position $(r_p, \phi_p, z_p)$ of the potential probe, 
the pipette is mounted onto a three-axis traversing robot. 
The $\phi_p$-axis is required to make measurements even 'behind' the
immersed ring (3) by rotating the probe between 0$^\circ$ and 180$^\circ$. 
In order not to  deform or crash the probe, a safety distance of 1mm
to the vessel and to the ring has been chosen. 
The extension of the scan-region (7), as well as the spatial resolution (6) 
are the same for all arrangements  to get comparable results. In the 
present case a step size of $\delta r=\delta z=1.25$ mm is applied which leads to a 
raster-size of $65\times 33$ pixels. Thus a measurement time of 
about 2 hours is necessary to traverse through the whole scan region.
The eddy current signal is logged pointwise
to a PC which also controls the
traversing robot and the waveform generator.

% ################################
% posting
The final eddy current distribution is determined in the post-processing. 
In the  case of harmonic excitation the measured voltages 
$U_{12}(r_p,z_p,t)$ are fitted by 
a sine function with the amplitude  $a_r(r_p,z_p)$ and the 
phase $ \phi_r(r_p,z_p)$ (see \fref{i_u_harm_fig}).
For transient excitation, time-slicing through the acquired data results in 
$u_s(r_p,z_p)$ for each instant (see \fref{i_u_trans_fig}). Both fields $a_r,u_s$
are converted from voltage to current by equation \eref{u_i_eqn}.
To illustrate and compare the measured versus calculated distributions the 
computation of isolines was performed. The required software was developed
under usage of MATLAB (Mathworks Inc.).

\section{Results}

In this section we compare the measured eddy currents
with the analytical results (in the case of homogeneous fluid)
and with numerical results (in the case of an immersed torus).
It should be noted that the specified lengths in $r,z$ direction 
refers to $r_p,z_p$ (cp. figure 2). Hence, the shown r-axis is 
located 1 mm 
above the boundary of the half space.

% ... Alternative TGC
% ... -> größerer Dynamikbereich

\subsection{Homogeneous fluid}
We start with the case of a homogeneous fluid for which 
analytical solutions for the harmonic and transient 
excitation were given in section 2.

\subsubsection{Harmonic excitation}
In the case of harmonic oscillation we show both
the amplitude and the phase shift of the induced currents
for a frequency of 480 Hz (figure 5).

\begin{figure}[ht]
\centering
\includegraphics{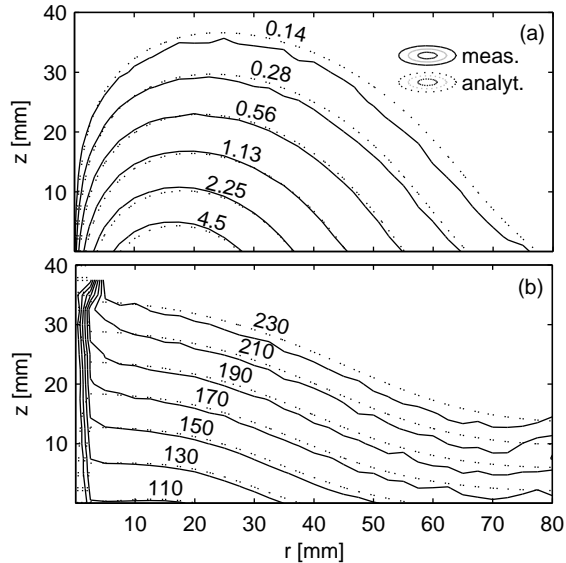}
\caption{Homogeneous fluid: contour lines of amplitude [A/cm$^2$] (a) and
phase-angle [$^\circ$] (b) of the azimuthal eddy current in the case of harmonic 
excitation with a frequency of 480 Hz and an excitation current of 1 A.}
\label{fig5}
\end{figure}

In addition to the measured values, we show also the results of 
equation (1).
Actually, we have also computed the eddy current
by the commercial FEM code OPERA. However,
in the adopted spatial resolution
the results are indistinguishable from the analytical ones.
In general, we see a good correspondence of
experimental and theoretical data
which expectedly gets worse for smaller radii where the
finite distance (6 mm) of the two electrodes of the potential
probe makes a precise measurement of the azimuthal currents
impossible.

% total-current-comparison

\subsubsection{Transient excitation}

In figure 6 we present the measured and
the theoretical eddy current distribution
for the case of pulsed excitation for the
time instants 200, 600, 1000 and 1400 $\mu$s.
These instants were already indicated as circles in the lower
panel of  figure 4.
Again, we observe a good correspondence,  despite the
used approximations with respect to geometry
and to the pulse shape.

\begin{figure}[ht]
\centering
\includegraphics{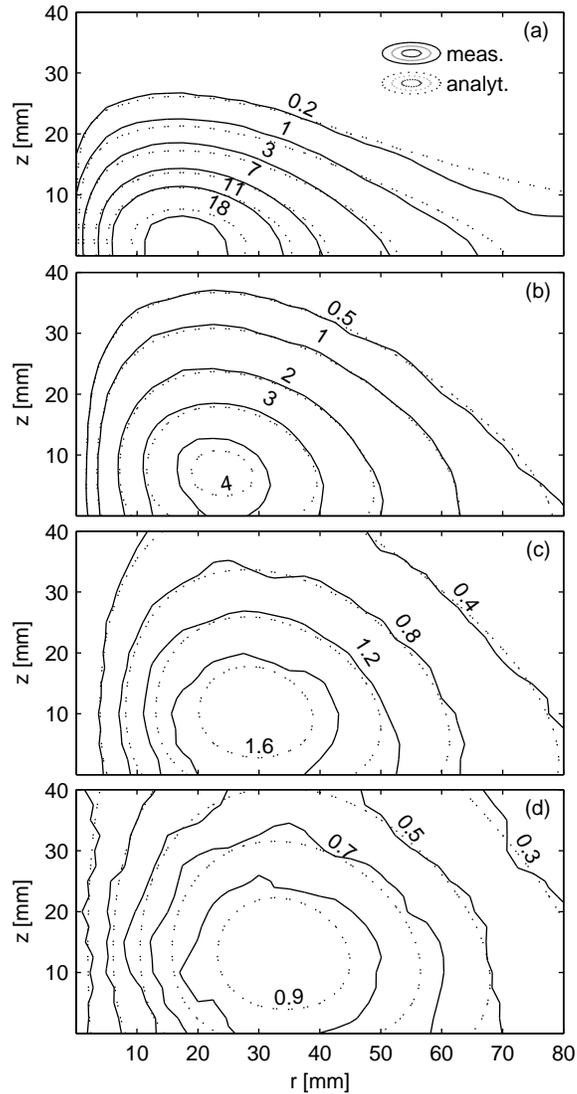}
\caption{Contour-plot of the azimuthal eddy 
current-distribution [A/cm$^2$] in the case of pulsed excitation with an
excitation current of 5 A. 
(a), (b), (c), (d) correspond to time instants,
200, 600, 1000, and 1400  $\mu$s, respectively. 
which were marked in \fref{i_u_trans_fig}.}
\label{fig6}
\end{figure}

It might also be instructive to compare the time evolution of the
total currents which is the integral in $r$ and $z$ direction of the
eddy current density (figure 7). In a wide range of the time evolution,
the misfit between the two data is of the order of a few percent.
Only at the very beginning and at the end (when the measured signal is
already very small), we get a discrepancy of around 10 per cent.

\begin{figure}[ht]
\centering
\includegraphics{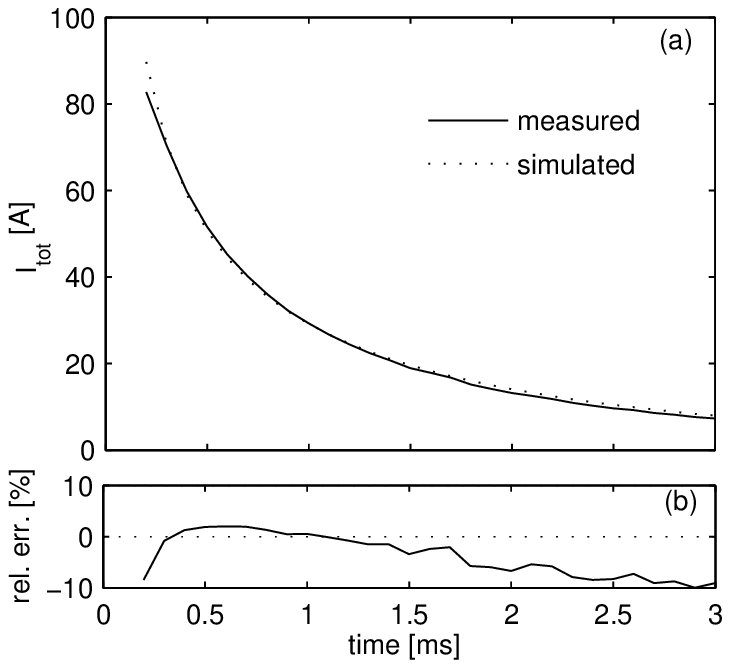}
\caption{Time evolution of the total current for the pulsed
current excitation in the case of a homogeneous fluid.}
\label{fig7}
\end{figure}

\subsection{Fluid with an immersed torus}
In order to study a slightly more complicated problem which
is not solvable anymore by an analytical expression,
we have immersed a PVC torus into the liquid metal.
This problem has
been solved again by means of the commercial
FEM solver OPERA which, in the case of a homogeneous fluid, had provided
results more or less identical to the analytical one. Note that
a similar transient eddy current
analysis was recently published \cite{TSUBOI}.

\subsubsection{Harmonic excitation}

In figure 8, we show again
the amplitude and the phase shift of the induced currents
for a excitation frequency of 480 Hz. The stronger
deviation to the case of a homogeneous fluid are
detected for the phase, while the amplitude is rather
unaffected by the insertion of the torus.

\begin{figure}[ht]
\centering
\includegraphics{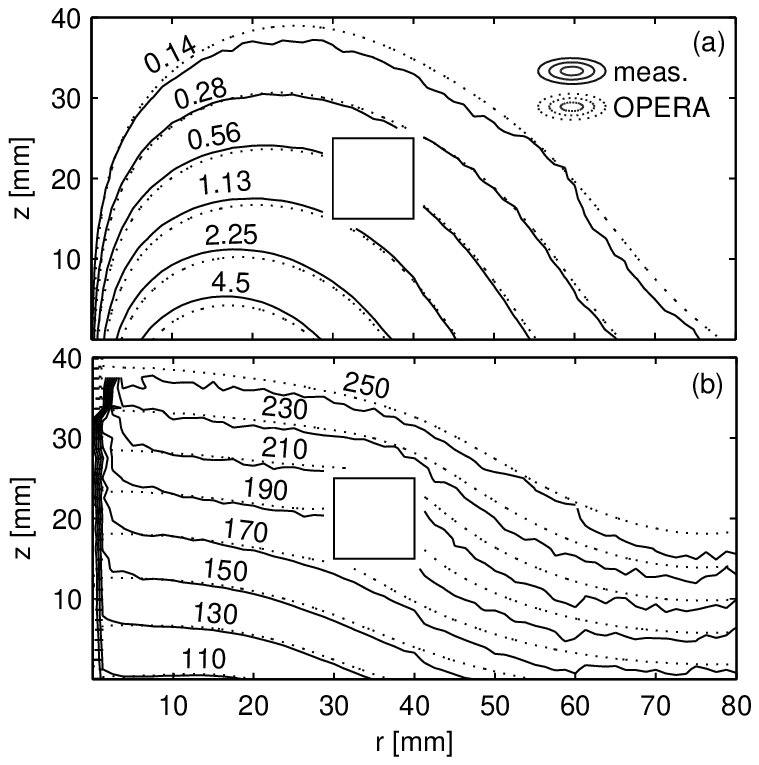}
\caption{Inhomogeneous fluid: contour lines of amplitude [A/cm$^2$] (a) and
phase-angle [$^\circ$] (b) of the azimuthal eddy current in the case of harmonic 
excitation with a frequency of 480 Hz and an excitation current of 1 A.}
\label{fig8}
\end{figure}

\subsubsection{Transient excitation}

In figure 9 we present again the measured and computed
eddy current distribution
for the
case of pulsed excitation for the time instants
200, 600, 1000 and 1400 $\mu$s.

\begin{figure}[ht]
\centering
\includegraphics{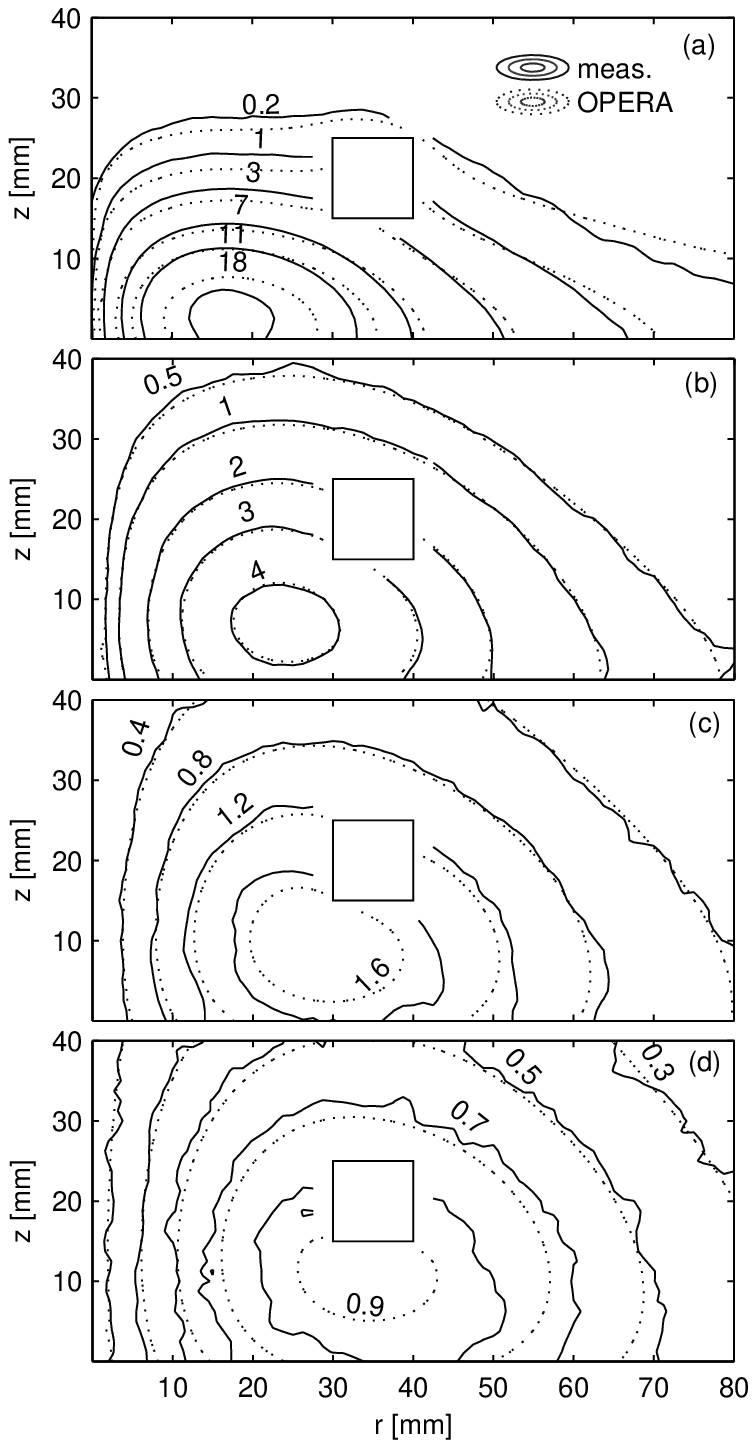}
\caption{Case of an inserted PVC ring with quadratic cross-section:
contour lines of the amplitude of the
azimuthal eddy current in the case of pulsed excitation with a
excitation current of 5 A. The full lines
are the measured data, the dotted lines are the numerical ones.
(a), (b), (c), (d) correspond to time instants,
200,600, 1000, and 1400  $\mu$s, respectively. }
\label{fig9}
\end{figure}

The deviation of the measured  total current from the numerically
obtained one (figure 10) is a bit larger in this case, which might be
due to the technical problems to measure the current close the
inserted PVC ring.

\begin{figure}[ht]
\centering
\includegraphics{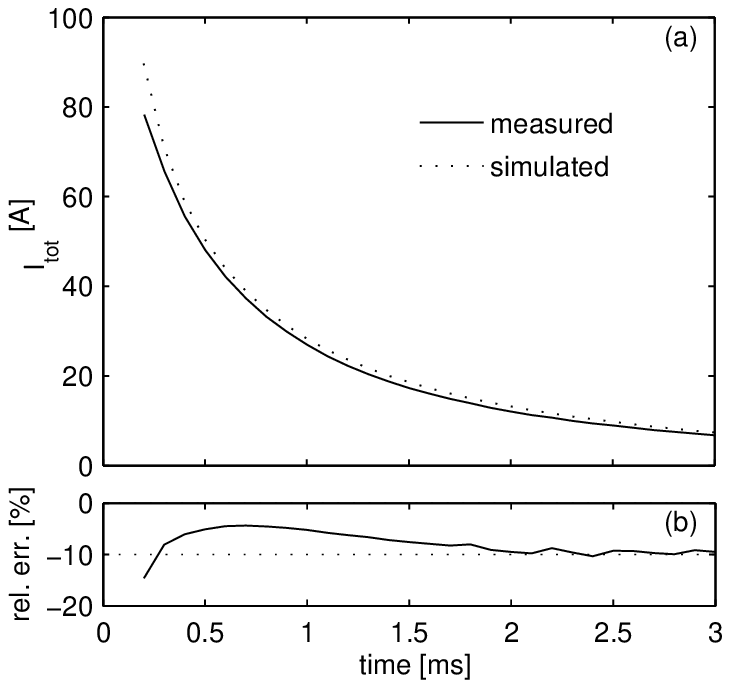}
\caption{Time evolution of the total current for the
pulsed current excitation
in the case with an immersed PVC ring.}
\label{fig10}
\end{figure}

\section{Conclusions}
We have measured the two-dimensional, axi-symmetric
distribution of eddy currents
in a liquid metal arising from harmonic or pulsed excitations
in a nearby coil. In both cases, the measured values
show a satisfactory
agreement with the analytical solution of
the corresponding problem for a conducting half-space.
Expectedly, a  non-conducting torus immersed into the fluid
disturbs the eddy current distribution, which was 
also confirmed  by means of a commercial numerical solver (OPERA).

% especially 
% motion induced currents ...

% we assume that there is no motion is induced by the excitation

Although we have restricted our
interest to axi-symmetric problems in which
the only relevant current component is the azimuthal one,
the
method can easily be generalized to fully three-dimensional
scans of all three
eddy current components. This way, it
might help to validate numerical models
for non-destructive testing and magnetic
inductance tomography.

\ack
This work was supported by Deutsche Forschungsgemeinschaft
in frame of the Sonderforschungsbereich 609.


\section*{References}
\begin{thebibliography}{10}
\bibitem{GRIFFITH} Griffith H 2001 Magnetic induction tomography {\it Meas. Sci. Technol.} {\bf 12}
1126-31
\bibitem{BLITZ}Blitz J 1997 {\it Electrical and Magnetic Methods of Non-destructive Testing} {Springer: Berlin }
\bibitem{MST}Stefani F and Gerbeth G 2000 A contactless method for velocity reconstruction in 
electrically conducting fluids {\it Meas. Sci. Technol.} {\bf 11} 758-65
\bibitem{CIFT}Stefani F, Gundrum T and Gerbeth G 2004 Contactless inductive flow tomography
 {\it Phys. Rev. E} {\bf  70}  056306
 \bibitem{PRIEDE}Priede J, Buchenau D and Gerbeth  G 2006 Contactless electromagnetic
 induction flowmeter based on phase shift measurements {\it Proc. 5th Int. Symp.
 Electromagnetic Processing of Materials (EPM2006)} (ISIJ, Tokyo) p~735
\bibitem{THESS}Thess A, Votyakov E V and Kolesnikov Y 2006 Lorentz force
velocimetry  {\it Phys. Rev. Lett.} {\bf 96} 164501
\bibitem{FUJIWARA} Fujiware K and Nakata T 1990 {\it COMPEL} {\bf 9} 137
\bibitem{FOERSTER} F\"orster F 1996 {\it Nondestructive Testing Handbook} Vol. 4, 2nd edn, {American 
Society for Nondestructive Testing: Columbus OH}, {sections 4 and 5}
\bibitem{BLITZALAGOA}Blitz J and Alagoa K D 1985 Eddy-current testing of Wood's
metal models for inclined cracks {\it NDT Intern.} {\bf 18} 269-73 
\bibitem{SMITH} Smith RA and Hugo G R 2001 {\it Insight} {\bf 41} 14-25
\bibitem{EDDYBUCH} Tegopoulos J A and Kriezis E E 1985 {\it Eddy Currents in Linear Conducting
Media} {Elsevier: Amsterdam}
\bibitem{TSUBOI}Tsuboi H, Seshima N, Sebestyen I, Pavo J, Gyimothy S, and Gasparics A 2004
{\it IEEE Trans. Magn.} {\bf 40} 1330-1333
\end{thebibliography}
\end{document}